\documentclass[11pt,twoside]{article}
\usepackage[pdftex]{graphicx}
\usepackage{amsmath}
\usepackage{amssymb}
\usepackage{cite}

\usepackage{subfig}
\usepackage{tabularx}
\usepackage{dcolumn}% Align table columns on decimal point
\usepackage{braket}
\usepackage[justification=raggedright]{caption}

\DeclareMathOperator{\Tr}{Tr}

\graphicspath{{../}}

\begin{document}
\newcommand{\pst}{\hspace*{1.5em}}

\newcommand{\rigmark}{\em Journal of Russian Laser Research}
\newcommand{\lemark}{\em Volume 30, Number 5, 2009}

%\lhead[\fancyplain{\rigmark, {\em \lemark}}{\rigmark}]{\fancyplain{\rigmark, {\em \lemark}}{\lemark}}
%\chead{}\rhead[\fancyplain{}{\lemark}]{\fancyplain{}{\rigmark}}
%\plainfootrulewidth 0.4pt
\newcommand{\be}{\begin{equation}}
\newcommand{\ee}{\end{equation}}
\newcommand{\bm}{\boldmath}
\newcommand{\ds}{\displaystyle}
\newcommand{\bea}{\begin{eqnarray}}
\newcommand{\eea}{\end{eqnarray}}
\newcommand{\ba}{\begin{array}}
\newcommand{\ea}{\end{array}}
\newcommand{\arcsinh}{\mathop{\rm arcsinh}\nolimits}
\newcommand{\arctanh}{\mathop{\rm arctanh}\nolimits}
\newcommand{\bc}{\begin{center}}
\newcommand{\ec}{\end{center}}

\thispagestyle{plain}

\label{sh}

%\lfoot[\fancyplain{\ \\[1mm] \thepage}{\ \\[1mm]\thepage}]{\fancyplain{}{}}

\begin{center} {\Large \bf
\begin{tabular}{c}
Deriving Entropic Inequalities for Two Coupled Superconducting Circuits
\end{tabular}
 } \end{center}

\bigskip

\bigskip

\begin{center} {\bf
Evgenii Glushkov$^{1,2*}$, Anastasiia Glushkova$^1$  and V. I. Man'ko$^{1,3}$
}\end{center}

\medskip

\begin{center}
{\it
$^1$Moscow Institute of Physics and Technology\\
Institutskii per. 9, Dolgoprudnii, Moscow Region 141700, Russia

\smallskip

$^2$National University of Science and Technology "MISiS"\\
Leninskii pr., 4, Moscow 119991, Russia 

\smallskip

$^3$Lebedev Physical Institute, Russian Academy of Sciences\\
Leninskii pr., 53, Moscow 119991, Russia 
}
\smallskip

$^*$Corresponding author e-mail:~~eugene.glushkov@gmail.com\\
\end{center}

\begin{abstract}\noindent
We discuss the known construction of two interacting superconducting circuits, based on Josephson junctions, that can be precisely engineered and easily controlled. In particular, we use the parametric excitation of two circuits, realized by an instant change of the qubit coupling, to study entropic and information properties of the density matrix of the composite system. The density matrix is obtained from the initial thermal state and is then analyzed in the approximation of small perturbation parameter and sufficiently low temperature. We also check the subadditivity condition for this system both for von Neumann and deformed entropies and look at the dependance of mutual information on the temperature of the system. Finally, we discuss the applicability of this approach to describe such system of two coupled superconducting qubits as harmonic oscillators with limited Hilbert space.
\end{abstract}

\medskip


\section{Introduction}
The idea to use the Josephson junction to engineer superconducting circuits, where current and voltage are considered analogs of position and momentum in a parametric oscillator \cite{dodonov1989correlated}, is now employed to study the properties of qubits, associated with the state of such circuits. The non-stationary quantum states of current and voltage in such circuits have been extensively studied \cite{manko1994correlated,kiktenko2015multilevel,dodonov2008photon,dodonov2009photon} for more than two decades and enormous progress in experimental realization \cite{steffen2006state,shalibo2012quantum} of these systems has been made, which gave rise to the whole new area of research called quantum information processing. The recent works \cite{fujii2011quantum,dodonov2014analytical,veloso2015prospects,monteiro2016anti} study the information properties of qubits as the resource for future quantum technologies. The particular properties of the multi-qubit states to be analyzed are entropic and information inequalities \cite{shannon1948note, holevo2012quantum}, which serve as the basis for quantum information processing. 

In this work we aim at studying the subadditivity condition for von Neumann and q-entropies of the bipartite quantum system \cite{chernega2014tomographic, chernega2015deformed, man2014separability,man2014entanglement,man2014quantum,chernega2015no}, on the specific example of two coupled superconducting circuits \cite{glushkov2015testing}. To study entropic and information properties of the composite system we use the parametric excitation of two circuits, realized by an instant change of the qubit coupling. Initially the system is considered to be in a thermal state with a corresponding thermal density matrix, which is then analyzed in the approximation of small perturbation parameter and sufficiently low temperature. We also look at the dependance of mutual information on temperature of the system and discuss the applicability of our approach to describe two coupled superconducting qubits as harmonic oscillators with limited Hilbert space.

\section{Theoretical model}
We start with a simple model of two interacting harmonic oscillators  in a thermal state. As we would like to consider only the first two levels in each of the circuits, we are working in the limit $T \to 0$. This approximation and its applicability will be more thoroughly discussed in Section \ref{sec:applicability}.

\begin{figure}[h]
	\centering
	\includegraphics[width=0.6\columnwidth]{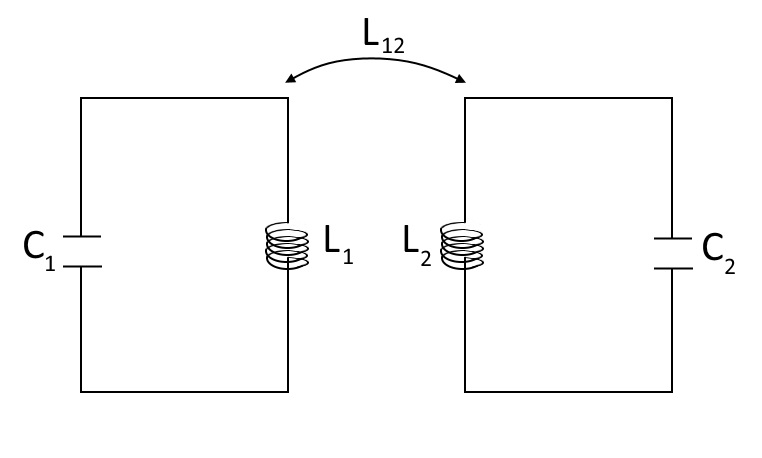}
	\caption{Schematic representation of two superconducting circuits, modelled as harmonic oscillators, coupled by a mutual inductance.}
	\label{fig:fig1}
\end{figure}

The Hamiltonian of the system of two coupled resonant circuits, depicted in the Fig. \ref{fig:fig1} can be written in the following simple form:

\begin{equation}\label{eq:hamiltonian}
	\hat{H}=\hat{H}_{1}+\hat{H}_{2}+\hat{V},
\end{equation}

where first two terms correspond to two independent LC-circuits and the third term defines coupling between them:

\[ 
	\hat{H}_{i}=\frac{L_{i}\hat{I}_{i}^2}{2} + \frac{\hat{Q}_{i}^2}{2C_{i}}, ~~\hat{V}=L_{12}\hat{I}_{1}\hat{I}_{2},
	~~i = 1,2
\]

For the convenience and due to the duality between mechanical oscillators and LC-circuits, we now introduce canonical position and momentum operators, which we will be using hereinafter, through the following change of variables \cite{fedorov2015tomographic}:

\[
	\hat{x}_j = -L_j C^{1/2} \hat{I}_j, 
	~~\hat{p}_j = C_j^{-1/2} \hat{Q}_j, 
	~~\left[\hat{x}_j, \hat{p}_k \right] = i\hbar \delta_{jk},
	~~ j, k = 1,2
\]

\noindent Using this substitution we can rewrite Eq. \ref{eq:hamiltonian} in the following form:
\begin{equation}
	\hat{H}=\frac{\hat{p}_{1}^2}{2m}+\frac{m\omega_{1}^2\hat{x}_{1}^2}{2}+\frac{\hat{p}_{2}^2}{2m}+\frac{m\omega_{2}^2\hat{x}_{2}^2}{2}+gm\hat{x}_{1}\hat{x}_{2}\omega_{1}\omega_{2}
\end{equation}

\noindent where $g = L_{12}(L_1 L_2)^{-1/2}$ is the qubit coupling constant and $\omega_{1,2}$ are the eigenfrequencies of the circuits. For simplicity, here and later we assume that $m = 1$, $\omega_1 = 1$,  $\omega_2 = \lambda$. 

Next we would like to get rid of the cross term and diagonalize the Hamiltonian \cite{abdalla1994quantum}. To do that we apply the following rotation by an angle $\phi$, using the theorem that a quadratic form is reduced to a diagonal form by an orthogonal transformation:
\begin{equation}
	\begin{pmatrix}
		x_1 \\
		x_2 \\         
	\end{pmatrix}
	=  \begin{pmatrix}
		\cos\phi & \sin\phi \\
		-\sin\phi & \cos\phi \\         
	\end{pmatrix}
	\begin{pmatrix}
		x_1' \\
		x_2' \\         
	\end{pmatrix}
\end{equation}

\noindent After this transformation, we obtain the following Hamiltonian: 
\begin{multline}\label{eq:expanded_H}
\hat{H}=\frac{\hat{p}_{1}^2}{2}+\frac{{\hat{p}}_{2}^2}{2}+
\frac{\cos^2\phi\hat{x}_{1}'+\sin^2\phi\hat{x}_{2}'^2}{2}+
\cos\phi\sin\phi\hat{x}_{2}'\hat{x}_{1}'+\\
+\frac{\lambda^2(\sin^2\phi\hat{x}_{1}'+\cos^2\phi\hat{x}_{2}')}{2}
-\lambda^2\sin\phi\cos\phi\hat{x}_{2}'\hat{x}_{1}'+\\
+g\lambda(-\cos\phi\sin\phi\hat{x}_{1}'^2+\cos\phi\sin\phi\hat{x}_{2}'^2)+g\lambda(\cos^2\phi-\sin^2\phi)\hat{x}_{2}'\hat{x}_{1}'
\end{multline}

\noindent Equating to zero the cross terms from Eq. \ref{eq:expanded_H}, we obtain the system of two independent harmonic oscillators:
\begin{equation}
\hat{H}=\frac{\hat{p'}_{1}^2}{2}+\frac{\Omega_{1}^2\hat{x'}_{1}^2}{2}+\frac{\hat{p'}_{2}^2}{2}+\frac{\Omega_{2}^2\hat{x'}_{2}^2}{2},
\end{equation}

\noindent where the new resonant frequencies and the rotation angle in the approximation of small angles are:
\begin{equation}\label{eq:new_freqs}
\Omega_1^2 \approx  1 - 2g\lambda\phi+\lambda^2\phi^2,
~~\Omega_2^2 \approx \phi^2 + \lambda^2 + 2g\lambda\phi,
~~\phi\approx g\lambda/(\lambda^2-1)~.
\end{equation}

\section{Density Matrix}

As our oscillators are in the thermal state in the low temperature limit the density matrix of the system, reduced to the $4 \times 4$ subspace, will take the following form: 

\begin{equation}
	\rho = 
	\frac{1}{Z(T)}
	\begin{pmatrix}
		\exp^{-\frac{E_{00}}{T}} & 0 & 0 & 0 \\
		0 & \exp^{-\frac{E_{01}}{T}} & 0 & 0 \\
		0 & 0 & \exp^{-\frac{E_{10}}{T}} & 0 \\
		0 & 0 & 0 & \exp^{-\frac{E_{11}}{T}} \\
	\end{pmatrix},
\end{equation}
\vspace{0.1cm}

\noindent where $E_{nm} = \Omega_1 (n+1/2)+\Omega_2 (m+1/2) $ and $Z(T) = Z_1(T) \cdot Z_2(T) = \left(4 \sinh{\frac{\omega_1}{2T}}\sinh{\frac{\omega_2}{2T}}\right)^{-1}$.

However, if we want to return to our initial system with two coupled resonant circuits, we have to perform transformation, which decomposes the old eigenstates in the new rotated basis:

\begin{equation}\label{eq:trans}
\tilde{\rho} = U^{-1} \rho~ U
\end{equation}

In the basis of the harmonic oscillators eigenstates we denote the coefficients of the transformation $U_{nmn'm'}$ and calculate them using eigenfunctions of the harmonic oscillators:
\begin{align}
	U_{nmn'm'} = \frac{1}{\pi} 
	\int_{-\infty}^{\infty}
	\frac{e^{-\frac{x_1^2}{2} - \frac{x_2^2}{2l^2} -\frac{x_1^{\prime 2}}{2L_1^2} - \frac{x_2^{\prime 2}}{2L_2^2}} }{\sqrt{lL_1L_2} \sqrt{2^{n}2^{m} n! m!}}
	\frac{dx_1dx_2}{\sqrt{2^{n'}2^{m'} n'! m'!}}  \cdot 
	H_{n} \left( x_1 \right) 
	H_{m} \left( \frac{x_2}{l} \right)
	H_{n'} \left(\frac{x'_1}{L_1}\right) 
	H_{m'} \left(\frac{x'_2}{L_2}\right),
\end{align}
\noindent where:
\begin{align}
	l = \sqrt{\frac{\hbar}{m\omega}}=\sqrt{\frac{1}{\lambda}}, \omega = \lambda, L_1 = \sqrt{\frac{1}{\Omega_1}}, L_2 = \sqrt{\frac{1}{\Omega_2}} \nonumber
\end{align}

\noindent and $H_i$ are the corresponding Hermite polynomials.

\begin{figure}[ht]
	\centering
	\subfloat[$T = 100$ mK]{
		\includegraphics[width=.33\columnwidth]{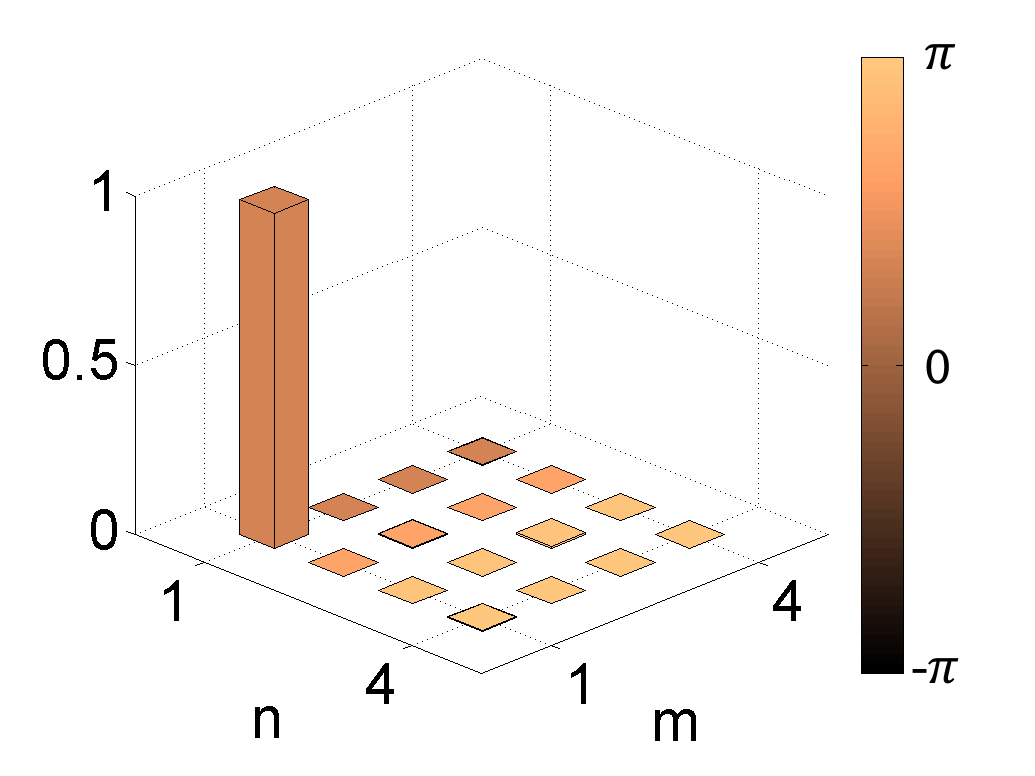}}
	\subfloat[$T = 250$ mK]{
		\includegraphics[width=.33\columnwidth]{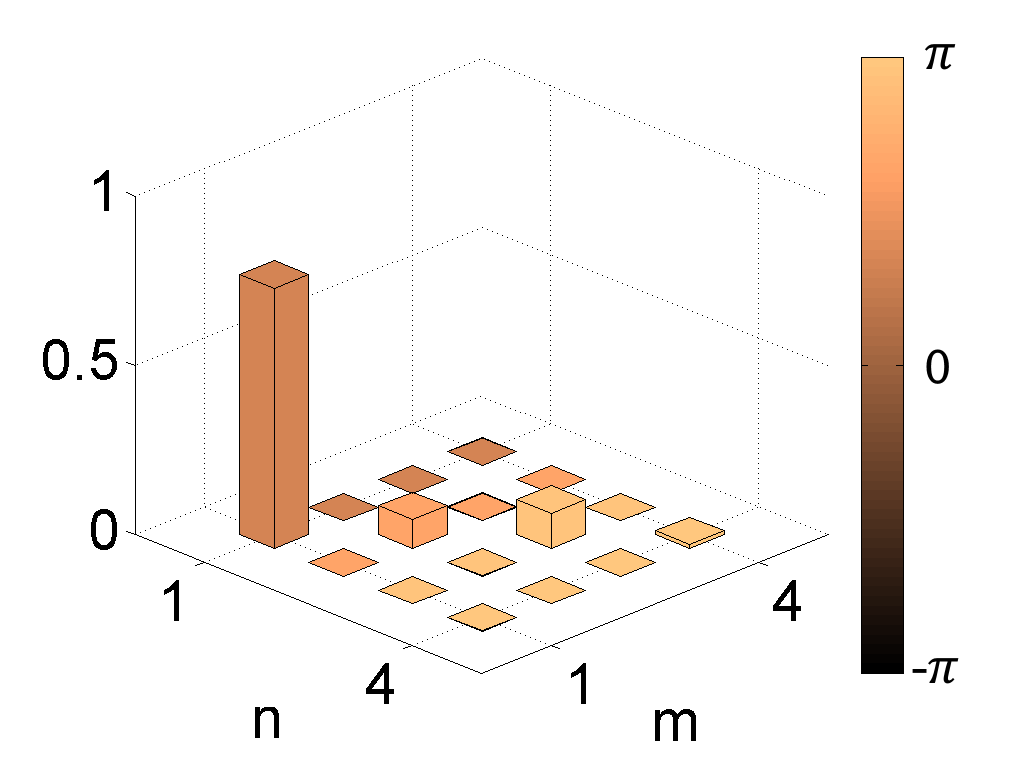}}
	\subfloat[$T = 500$ mK]{
		\includegraphics[width=.33\columnwidth]{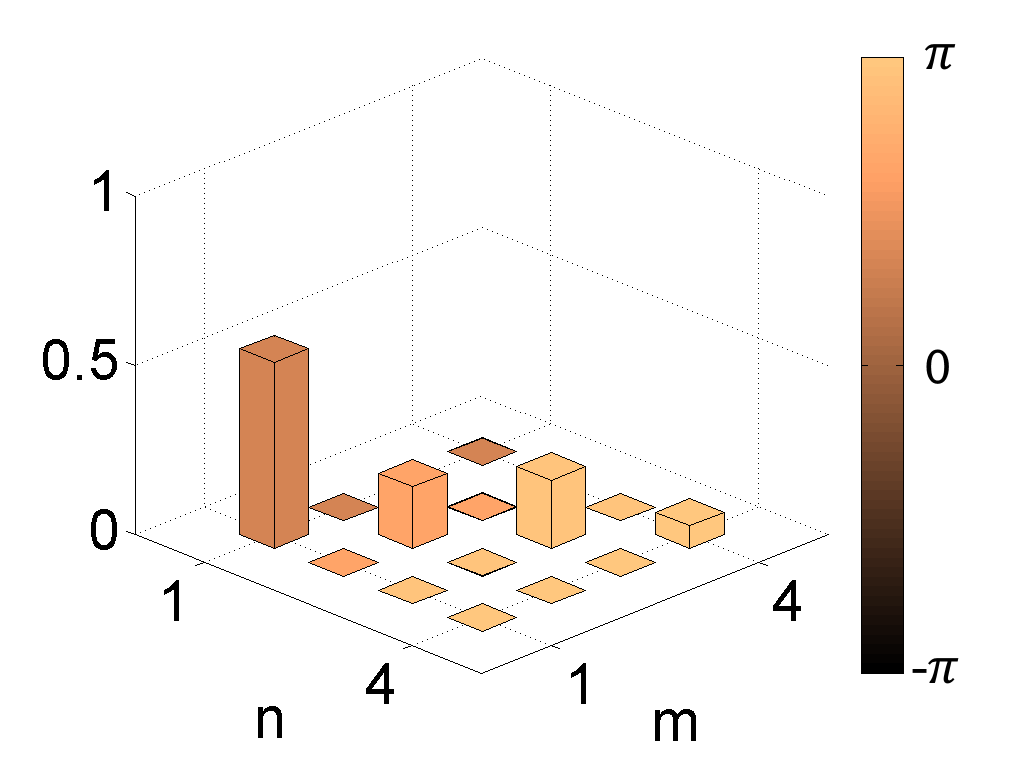}}
	\caption{(color online) Density matrices, calculated from Eq. \eqref{eq:trans}, at different temperatures. Off-diagonal elements are nonzero, but not visible due to the scale difference.} 
	\label{fig:DM}
\end{figure}

After calculating the matrix elements of the transformation $U_{nmn'm'}$ (see Appendix for the details) we obtain the final density matrix to work with from the Eq. \ref{eq:trans}, which can be seen in the Fig. \ref{fig:DM} at various temperatures of the system. One can see, that the approximation used is valid only for very low temperatures, below 100 mK, which is standard for superconducting qubits. At higher temperatures the populations of the higher energy levels become non-negligible, thus demanding to take into account larger subspace, than the $4\times4$ matrix we are discussing here.

\section{\label{sec:theory}Calculating entropy}

Using the obtained density matrix, we can calculate the entropical properties of the system. The main properties we will be looking at are entropies of the bipartite system and its subsystems (single qubits) and the mutual  information, which can be deduced from the subadditivity condition. 

\noindent The two types of entropies we are discussing in this work are the \textbf{von Neumann} entropy ($\label{eq:neumann}S = - Tr \tilde{\rho} \ln \tilde{\rho}$) and the \textbf{Tsallis} entropy, which equals the von Neumann entropy in the limit of $q \rightarrow 1$:
\begin{equation}\label{eq:tsallis}
	S_q^{T} = - Tr \tilde{\rho} \ln_{q} \tilde{\rho} ,
\end{equation}

\noindent where the q-logarithm is defined as follows: 

\begin{equation}\label{qentr}
	\ln_{q>0}\rho = 
	\begin{cases}\frac{\rho^{q - 1} - \hat{I}}{1 - q} ,q \neq 1\\
		\ln \rho, q = 1
	\end{cases}\nonumber
\end{equation}

\begin{figure}[h]
	\centering
	\includegraphics[width=1.0\columnwidth]{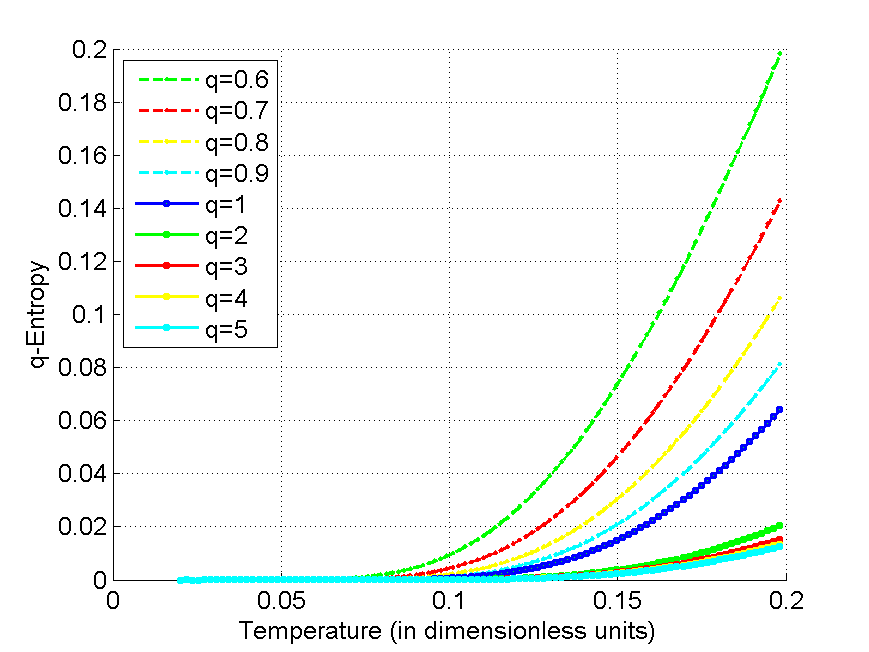}
	\caption{(color online) Dependance of the q-entropies on temperature calculated from Eq. \ref{eq:tsallis}.}
	\label{fig:entropies}
\end{figure}

As the matrix elements of the density matrix depend on temperature of the system, we calculated the Eq. \ref{eq:tsallis} for various temperatures and displayed the results of these calculations in Fig. \ref{fig:entropies}. The temperature here and later is chosen to be in units of frequency, relative to the eigenfrequency of the qubits $\omega$. 

One can see, that the q-entropies in Fig. \ref{fig:entropies} are grouped in such a way, that the entropies with $q < 1$ lie higher than the von Neumann entropy ($q = 1$), while the entropies with $q > 1$ lie lower. The meaning of this behavior can be understood in the following way. The higher is the value of q, the higher is the influence of the biggest terms in the distribution on entropy and the more deterministic is the behavior of the system. As the main elements of our density matrix are the four diagonal ones, higher q-values lead to lower value of q-entropy for our matrix.

\section{\label{sec:subadd}Verifying subadditivity condition}
To check the subadditivity condition for the bipartite system, consisting of two resonant circuits, we divide the calculated density matrix from Eq. \ref{eq:trans} into density matrices of the subsystems as follows:

\begin{equation} \label{eq:rhotilde1}
		\tilde{\rho}_1 = 
		\begin{pmatrix}
		\rho_{11}+\rho_{22} & \rho_{13}+\rho_{24} \\
		\rho_{31}+\rho_{42} & \rho_{33}+\rho_{44} \\
		\end{pmatrix}, ~~
		\tilde{\rho}_2 =
		\begin{pmatrix}
		\rho_{11}+\rho_{33} & \rho_{12}+\rho_{34} \\
		\rho_{21}+\rho_{43} & \rho_{22}+\rho_{44} \\
		\end{pmatrix}
\end{equation}

\noindent And the subadditivity condition for the corresponding q-entropies reads:

\begin{equation} \label{eq:subad}
	S^T_q (\tilde{\rho}) \le  S^T_q (\tilde{\rho}_1) + S^T_q (\tilde{\rho}_2) 
\end{equation}

\noindent One can also deduce the value of mutual information out of it:

\begin{equation}\label{eq:mutual}
	I = S^T_q(\rho) - S^T_q(\rho_1) - S^T_q(\rho_2) \ge 0
\end{equation}

\vspace{-0.1cm}
\begin{figure}[h]
	\centering
	\includegraphics[width=1.0\columnwidth]{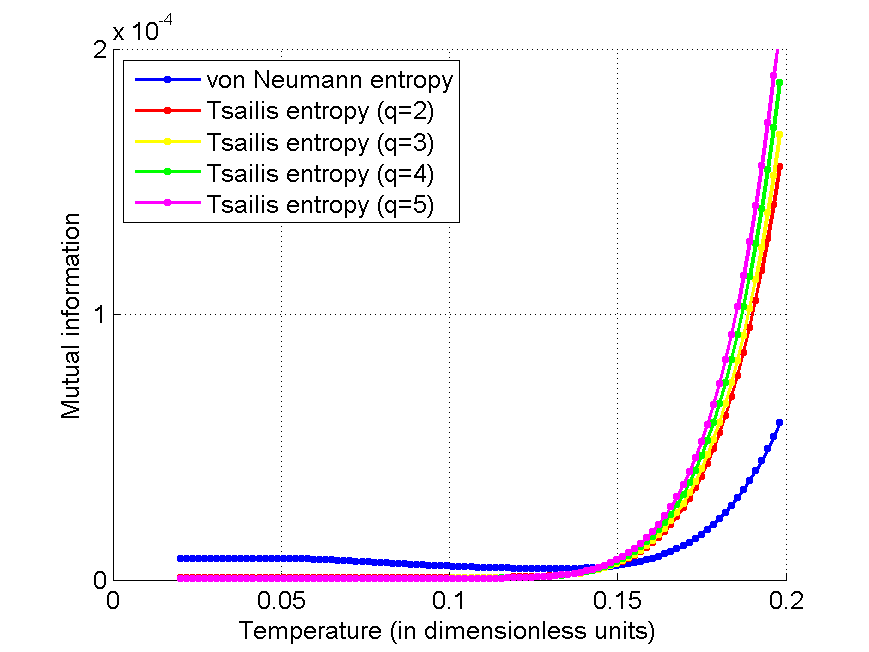}
	\caption{(color online) Mutual information related to temperature.}
	\label{fig:mutual}
\end{figure}

Using Eq. \ref{eq:mutual} we calculate the value of mutual information for various temperatures of the system and present them in Fig. \ref{fig:mutual}. The fact that the value of mutual information stays positive in the whole temperature range shows that the subadditivity condition holds for our system.

\section{Approximation applicability}\label{sec:applicability}
\begin{figure}[h]
	\centering
	\subfloat[Relation of purities.]{
		\label{fig:purities1}
		\includegraphics[width=0.5\columnwidth]{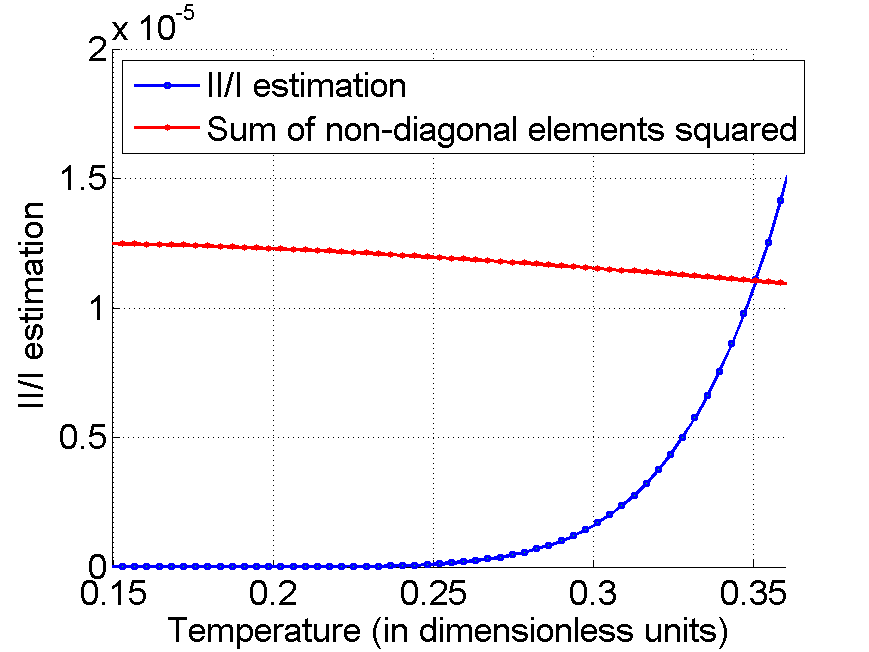}}
	\subfloat[Purity for different temperatures.]{
		\label{fig:purities2}
		\includegraphics[width=0.5\columnwidth]{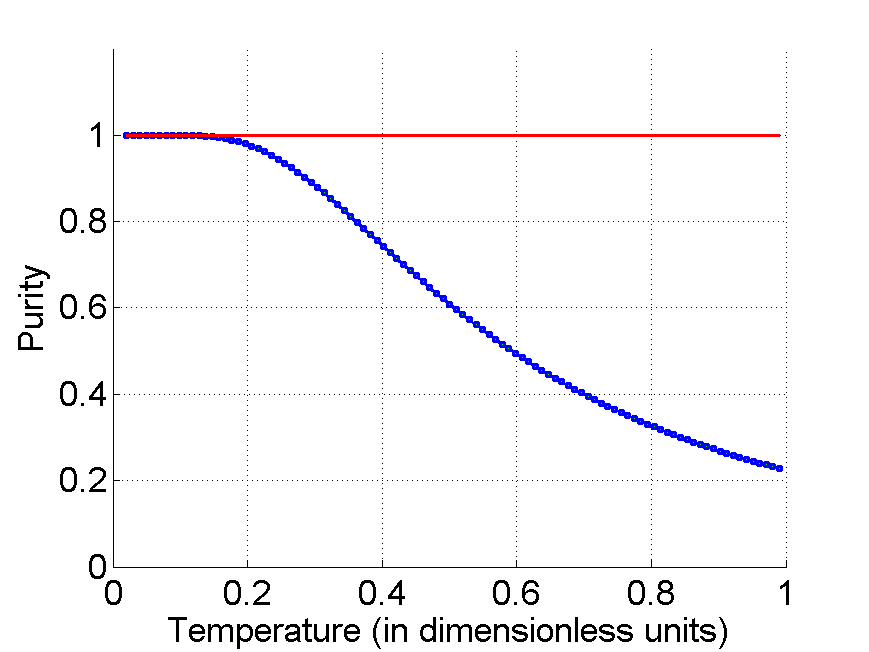}}
	\caption{(color online) Applicability limits. (a) The inverse relation of purity of the 4-level approximation to the purity of the remaining part of the density matrix. (b) The dependance of purity of the 4-level approximation on temperature.}
	\label{fig:purities}
\end{figure}

In this section we discuss the applicability of the approach we used to approximate the system of two harmonic oscillators by two qubits in the limit of low temperature. We look at the density matrix of our system and calculate the purity parameter $\mu = \Tr{\rho^2}$, first for the reduced $4\times 4$ matrix (I) and then for the remaining part of the density matrix (II). The result of these calculations is shown in  the Fig. \ref{fig:purities1} in relation to temperature of the system. We also plot the sum of non-diagonal elements of the reduced $4\times 4$ matrix to see the region, where it's bigger that the purity error.

Additionally in the Fig. \ref{fig:purities2} we plot the purity of the $4\times 4$ density matrix versus temperature to show the applicability of our approach. One can see that the $4\times 4$ subspace approximation is valid below $T \approx 0.2$, where $\mu \approx 1$. 

Based on these calculations, we believe that the approach used is valid for temperatures up to 100 mK, or 0.2 in relative units. So, for all the calculations with the reduced $4\times 4$ density matrix, we normalize its elements to get $\mu = 1$.

\section{Summary and conclusions}

We have analyzed the system of two coupled superconducting circuits, modeled as two interacting harmonic oscillators in the low-temperature limit, and derived its density matrix in the approximation of small perturbation. We have further checked that the calculated density matrix of the bipartite system satisfies the entropic inequalities both for the von Neumann and Tsailis entropy and looked at the dependance of mutual information on the temperature of the system. Finally, we evaluated the purity parameter of the system and verified the applicability of our approach to describe the system of two coupled superconducting qubits as harmonic oscillators with limited Hilbert space.

As the next step we would like to compare our findings with experimental data and generalize our approach for arbitrary rotation angle $\phi$ and for larger number of qubits, which we will do in future publications.

\section*{Acknowledgments}
The support by the Ministry for Education and Science of Russian Federation in the framework of Increased Competitiveness Program of the National University of Science and Technology MISIS under contract no. K2-2014-025 is gratefully acknowledged. We would also like to thank E. Kiktenko for stimulating discussions.

%==============================================================================
%==End of content==============================================================
%==============================================================================

%--References------------------------------------------------------------------

\bibliography{paper}

\section*{Appendix}\label{sec:appendix}

Here we describe in more detail how the matrix elements were calculated. In the basis of the eigenstates $\ket{m,n}$ we denote the coefficients of decomposition $U_{nmn'm'}$ and calculate the matrix elements of the $ \hat{U} $ in the eigenvalues basis of the Hamiltonian (Fock basis).

\noindent For the transition from an old basis to a new one $\ket{e_{k}} \to \ket{\tilde{e_{k}}}$ one can write the following decomposition: 
\begin{multline}
\ket{\tilde{e_{k}}}=\sum\limits_{m} U_{km}\ket{e_{m}}
\sum\limits_{m} \bra{e_{n}}U_{km}\ket{e_{m}}=\bra{e_{n}}\hat{1}\ket{\tilde{e_{k}}}=\\
=\sum\limits_{m} U_{km}\bra{e_{n}}\hat{1}\ket{e_{m}}=\sum\limits_{m} U_{km}\delta_{nm}=U_{kn}=\int_{-\infty}^{\infty} \Psi_{n}^*(x) \tilde{\Psi}_{k}(x) dx
\end{multline}

\noindent So, for the density matrix we obtain:
\begin{equation}
\rho_{nmn'm'} = \bra{nm}\hat{\rho}\ket{n'm'} = \frac{1}{Z(T)}\bra{nm}e^{-\beta\hat{H}}\ket{n'm'}
\end{equation}

\noindent Using the harmonic oscillator eigenfunctions: 
$\Psi_{n}(x)=\frac{1}{\sqrt{2^{n}n!\sqrt{\pi}}}e^{-\frac{1}{2}x^2}\hat{H}_n(x)$, where $\omega=1$, we get:

\begin{align}
U_{nmn'm'} &= \frac{1}{\pi} \int_{-\infty}^{\infty} 
\frac{e^{-\frac{x_1^2}{2} - \frac{x_2^2}{2l^2}} }{\sqrt{l}}
\frac{e^{-\frac{\Omega_1x_1^{\prime 2}}{2} - \frac{\Omega_2 x_2^{\prime 2}}{2}} }{\sqrt{L_1L_2}}
\frac{dx_1}{\sqrt{2^{n}2^{m} n! m!}}
\frac{dx_2}{\sqrt{2^{n'}2^{m'} n'! m'!}} \nonumber \\
& \cdot 
H_{n} \left( x_1 \right) 
H_{m} \left( \frac{x_2}{l} \right)
H_{n'} \left(\frac{x'_1}{L_1}\right) 
H_{m'} \left(\frac{x'_2}{L_2}\right)=\nonumber \\
& = \frac{1}{\pi} \int_{-\infty}^{\infty} 
dx_1 dx_2
\frac{e^{-\frac{x_1^2}{2} - \frac{x_2^2}{2l^2}} }{\sqrt{l}}
\frac{e^{-\frac{x_1^{\prime 2}}{2L_1^2} - \frac{x_2^{\prime 2}}{2L_2^2}} }{\sqrt{L_1L_2}}
\frac{1}{\sqrt{2^{n}2^{m} n! m!}}
\frac{1}{\sqrt{2^{n\prime}2^{m\prime} n'! m'!}} \nonumber \\
& \cdot 
H_{n} \left( x_1 \right) 
H_{m} \left( \frac{x_2}{l} \right)
H_{n'} \left( \frac{\cos\phi x_1+\sin\phi x_2}{L_1} \right) 
H_{m'} \left( \frac{-\sin\phi x_1+\cos\phi x_2}{L_2} \right) = \nonumber
\end{align}
\begin{align}
=& \frac{1}{\pi} \int_{-\infty}^{\infty}
\frac{e^{-\frac{x_1^2}{2} - \frac{x_2^2}{2l^2}} }{\sqrt{l}}
\cdot \frac{H_{n} \left( x_1 \right) }{\sqrt{2^{n'}2^{m'} n'! m'!}} 
\frac{H_{m} \left( \frac{x_2}{l} \right)}{\sqrt{2^{n}2^{m} n! m!}}
H_{n'} \left( \frac{\cos\phi x_1+\sin\phi x_2}{L_1} \right) 
H_{m'} \left( \frac{-\sin\phi x_1+\cos\phi x_2}{L_2} \right) \nonumber\\
& \frac{e^{-\frac{(\cos\phi x_1+\sin\phi x_2)^2}{2}(\lambda^2\sin^2\phi + \cos^2\phi - 2g\lambda\sin\phi\cos\phi) 
		- \frac{(-\sin\phi x_1+\cos\phi x_2)^2}{2}(\sin^2\phi + \lambda^2\cos^2\phi + 2g\lambda\sin\phi\cos\phi)} }{\sqrt{L_1L_2}}dx_1 dx_2 \nonumber
		\end{align}
		
		\noindent where $l = \sqrt{\frac{\hbar}{m\omega}}=\sqrt{\frac{1}{\lambda}} , \omega = \lambda, L_1 = \sqrt{\frac{1}{\Omega_1}}, L_2 = \sqrt{\frac{1}{\Omega_2}} \\
		\Omega_1^2 =\lambda^2\sin^2\phi + \cos^2\phi - 2g\lambda\sin\phi\cos\phi \\
		\Omega_2^2 =\sin^2\phi + \lambda^2\cos^2\phi + 2g\lambda\sin\phi\cos\phi \\
		$
		
\noindent In the approximation of small angles ($\cos\phi\approx1$ and $\sin\phi\approx\phi$) last terms are:
		\begin{align}
		&H_{n'} \left( \frac{\cos\phi x_1+\sin\phi x_2}{L_1} \right) 
		H_{m'} \left( \frac{-\sin\phi x_1+\cos\phi x_2}{L_2} \right) \approx H_{n'} \left( \frac{x_1 + \phi x_2}{L_1} \right) 
		H_{m'} \left( \frac{x_2 - \phi x_1}{L_2} \right) \nonumber\\
		&\Omega_1^2 \approx \lambda^2\phi^2 + 1 - 2g\lambda\phi\nonumber \\
		&\Omega_2^2 \approx \phi^2 + \lambda^2 + 2g\lambda\phi \nonumber
		\end{align}

Here we denote K = $\sqrt{lL_1L_2}\approx(\lambda^2(\lambda^2\phi^2 + 1 - 2g\lambda\phi)(\phi^2 + \lambda^2 + 2g\lambda\phi))^{-1/8} $
and use $H_0(x) = 1, H_1(x)=2x$. With these simplifications we obtain the following expression for the matrix elements:
		
		\begin{align}
		U_{nmn'm'} &= \frac{1}{\pi K} \int_{-\infty}^{\infty} 
		e^{-\left(\frac{1}{2}(\Omega_1+\phi^2\Omega_2+1)x_1^2 + (\phi\Omega_1-\phi\Omega_2)x_1x_2 +
			\frac{1}{2}(\phi^2\Omega_1+\Omega_2+\lambda)x_2^2 \right)
			}
			H_{n} \left( x_1 \right) 
			H_{m} \left( \frac{x_2}{l} \right)
			\nonumber \\ 
			& \cdot 
			H_{n'} \left(\frac{x_1+\phi x_2}{L_1}\right) 
			H_{m'} \left(\frac{x_2-\phi x_1}{L_2}\right)
			\frac{dx_1}{\sqrt{2^{n}2^{m} n! m!}}
			\frac{dx_2}{\sqrt{2^{n'}2^{m'} n'! m'!}} 
			\end{align}
			
\noindent Using this equation, we now calculate the matrix elements:
			\begin{flalign}
			U_{0000} & = \frac{1}{K\pi}
			\int_{-\infty}^{\infty} 
			e^{-\frac{x_1^2}{2} - \frac{x_2^2}{2l^2} - \frac{\Omega_1x_1^{\prime 2}}{2} - \frac{\Omega_2 x_2^{\prime 2}}{2}}dx_1 dx_2 &
			\end{flalign}
			
			\begin{flalign}
			U_{0001} & = \frac{\sqrt{2}(\phi^2 + \lambda^2 + 2g\lambda\phi)^{1/4}}{\pi K} 
			\int_{-\infty}^{\infty} (x_2 - \phi x_1)
			e^{-\frac{x_1^2}{2} - \frac{x_2^2}{2l^2} - \frac{\Omega_1x_1^{\prime 2}}{2} - \frac{\Omega_2 x_2^{\prime 2}}{2}}dx_1 dx_2 &
			\end{flalign}
			
			\begin{flalign}
			U_{0010} & = \frac{\sqrt{2}(\lambda^2\phi^2 + 1 - 2g\lambda\phi)^{1/4}}{\pi K}
			\int_{-\infty}^{\infty}  (x_1 + \phi x_2)
			e^{-\frac{x_1^2}{2} - \frac{x_2^2}{2l^2} - \frac{\Omega_1x_1^{\prime 2}}{2} - \frac{\Omega_2 x_2^{\prime 2}}{2} } dx_1 dx_2 &
			\end{flalign}
			
			\begin{flalign}
			U_{0011} & = \frac{2}{\pi \sqrt{\lambda} K^3}
			\int_{-\infty}^{\infty} (x_1+\phi x_2)(x_2-\phi x_1) 
			e^{-\frac{x_1^2}{2} - \frac{x_2^2}{2l^2} - \frac{\Omega_1x_1^{\prime 2}}{2} - \frac{\Omega_2 x_2^{\prime 2}}{2} } dx_1 dx_2 &
			\end{flalign}
			
			\begin{flalign}
			U_{0100} & = \frac{\sqrt{2\lambda}}{K\pi}  
			\int_{-\infty}^{\infty} x_2 
			e^{-\frac{x_1^2}{2} - \frac{x_2^2}{2l^2} - \frac{\Omega_1x_1^{\prime 2}}{2} - \frac{\Omega_2 x_2^{\prime 2}}{2} } dx_1 dx_2 &
			\end{flalign}
			
			\begin{flalign}
			U_{0101} &= \frac{2}{\pi K^3 (\lambda^2\phi^2 + 1 - 2g\lambda\phi)^{1/4}}  
			\int_{-\infty}^{\infty} x_2 (x_2-x_1\phi)
			e^{-\frac{x_1^2}{2} - \frac{x_2^2}{2l^2} - \frac{\Omega_1x_1^{\prime 2}}{2} - \frac{\Omega_2 x_2^{\prime 2}}{2} } dx_1 dx_2 &
			\end{flalign}
			
			\begin{flalign}
			U_{0110} & = \frac{2}{\pi K^3 (\phi^2 + \lambda^2 + 2g\lambda\phi)^{1/4}} 
			\int_{-\infty}^{\infty} x_2 (x_1+\phi x_2)
			e^{-\frac{x_1^2}{2} - \frac{x_2^2}{2l^2} - \frac{\Omega_1x_1^{\prime 2}}{2} - \frac{\Omega_2 x_2^{\prime 2}}{2} } dx_1 dx_2 &
			\end{flalign}
			
			\begin{flalign}
			U_{0111} & = \frac{2 \sqrt{2}}{\pi K^3}
			\int_{-\infty}^{\infty}  x_2 (x_2-\phi x_1)(x_1+\phi x_2)
			e^{-\frac{x_1^2}{2} - \frac{x_2^2}{2l^2} - \frac{\Omega_1x_1^{\prime 2}}{2} - \frac{\Omega_2 x_2^{\prime 2}}{2} } dx_1 dx_2 &
			\end{flalign}
			
			\begin{flalign}
			U_{1000} & = \frac{\sqrt{2}}{K\pi}  
			\int_{-\infty}^{\infty} x_1 
			e^{-\frac{x_1^2}{2} - \frac{x_2^2}{2l^2} - \frac{\Omega_1x_1^{\prime 2}}{2} - \frac{\Omega_2 x_2^{\prime 2}}{2} } dx_1 dx_2 &
			\end{flalign}
			
			\begin{flalign}
			U_{1001} &= \frac{2(\phi^2 + \lambda^2 + 2g\lambda\phi)^{1/4}}{K\pi} 
			\int_{-\infty}^{\infty} x_1 (x_2-\phi x_1) 
			e^{-\frac{x_1^2}{2} - \frac{x_2^2}{2l^2} - \frac{\Omega_1x_1^{\prime 2}}{2} - \frac{\Omega_2 x_2^{\prime 2}}{2} } dx_1 dx_2 &
			\end{flalign}
			
			\begin{flalign}
			U_{1010} & = \frac{2(\lambda^2\phi^2 + 1 - 2g\lambda\phi)^{1/4}}{K\pi} 
			\int_{-\infty}^{\infty} x_1 (x_1 + \phi x_2) 
			e^{-\frac{x_1^2}{2} - \frac{x_2^2}{2l^2} - \frac{\Omega_1x_1^{\prime 2}}{2} - \frac{\Omega_2 x_2^{\prime 2}}{2} } dx_1 dx_2 & 
			\end{flalign}
			
			\begin{flalign}
			U_{1011} & =  \frac{2\sqrt{2}}{\pi \sqrt{\lambda} K^3}
			\int_{-\infty}^{\infty} x_1 (x_2-\phi x_1)(x_1+\phi x_2)
			e^{-\frac{x_1^2}{2} - \frac{x_2^2}{2l^2} - \frac{\Omega_1x_1^{\prime 2}}{2} - \frac{\Omega_2 x_2^{\prime 2}}{2} } dx_1 dx_2 & 
			\end{flalign}
			
			\begin{flalign}
			U_{1100} & = \frac{2\sqrt{\lambda}}{K\pi}
			\int_{-\infty}^{\infty} x_1 x_2 
			e^{-\frac{x_1^2}{2} - \frac{x_2^2}{2l^2} - \frac{\Omega_1x_1^{\prime 2}}{2} - \frac{\Omega_2 x_2^{\prime 2}}{2} } dx_1 dx_2 & 
			\end{flalign}
			
			\begin{flalign}
			U_{1101} & = \frac{2\sqrt{2}}{\pi K^3 (\lambda^2\phi^2 + 1 - 2g\lambda\phi)^{1/4}}
			\int_{-\infty}^{\infty} x_1 x_2(x_2-\phi x_1) 
			e^{-\frac{x_1^2}{2} - \frac{x_2^2}{2l^2} - \frac{\Omega_1x_1^{\prime 2}}{2} - \frac{\Omega_2 x_2^{\prime 2}}{2} } dx_1 dx_2 & 
			\end{flalign}
			
			\begin{flalign}
			U_{1110} & = \frac{2\sqrt{2}}{\pi K^3 (\phi^2 + \lambda^2 + 2g\lambda\phi)^{1/4}}
			\int_{-\infty}^{\infty} x_1 x_2(x_1+\phi x_2) 
			e^{-\frac{x_1^2}{2} - \frac{x_2^2}{2l^2} - \frac{\Omega_1x_1^{\prime 2}}{2} - \frac{\Omega_2 x_2^{\prime 2}}{2} } dx_1 dx_2 & 
			\end{flalign}
			
			\begin{flalign}
			U_{1111} & = \frac{4}{\pi K^3}
			\int_{-\infty}^{\infty} x_1 x_2 (x_2-\phi x_1)(x_1+\phi x_2) 
			e^{-\frac{x_1^2}{2} - \frac{x_2^2}{2l^2} - \frac{\Omega_1x_1^{\prime 2}}{2} - \frac{\Omega_2 x_2^{\prime 2}}{2} } dx_1 dx_2 & 
			\end{flalign}
			
			\subsection*{Calculation of matrix elements}
			To calculate the integrals we use the following formula:
			\begin{equation}\label{integr}
			\int_{-\infty}^{\infty} e^{-(a_{11}x_1^2+a_{22}x_2^2+2a_{12}x_1 x_2) + b_1 x_1+b_2 x_2} dx_1 dx_2 =
			\frac{\pi}{\sqrt{det A}} e^{\frac{a_{22}b_1^2-2a_{12}b_1b_2+a_{11}b_2^2}{4 det A}} 
			\end{equation}
			
			\noindent where:
			\begin{align}
			a_{11} &= \frac{3}{2}+2\lambda^2\phi^2-2g\lambda\phi+\phi^2(\phi^2+2g\lambda\phi), \\
			a_{22} &=\frac{\lambda}{2}+\lambda^2+2\phi^2+2g\lambda\phi+ \phi^2(\lambda^2\phi^2-2g\lambda\phi), \\
			a_{12} &= a_{21} =  (\lambda^2\phi^2 + 1 - \lambda^2 - \phi^2 - 4g\lambda\phi)\phi, \\
			\det{A} &= (\frac{3}{2}+2\lambda^2\phi^2-2g\lambda\phi+\phi^2(\phi^2+2g\lambda\phi))
			(\frac{\lambda}{2}+\lambda^2+2\phi^2+2g\lambda\phi+ \phi^2(\lambda^2\phi^2-2g\lambda\phi)) - \nonumber\\
			&-(\lambda^2\phi^2 + 1 - \lambda^2 - \phi^2 - 4g\lambda\phi)^2\phi^2)
			\end{align}
			
			Then we obtain: $U_{0000} = \frac{1}{K\sqrt{\det{A}}}$
			
			To calculate the following integrals we differentiate Eq. \ref{integr} by the corresponding $a_{ij}$ :
			
			\begin{align}
			U_{0011} = \frac{(a_{12}(\phi^{2}-1) - \phi a_{22} + \phi a_{11} )}{(K\det{A})^{3/2}\sqrt{\lambda}}, ~~
			U_{0101} =  \frac{(\phi a_{12} + a_{11})}{K^3 (\det{A})^{3/2}(\lambda^2\phi^2 + 1 - 2g\lambda\phi)^{1/4}} 
			\\
			U_{0110} = \frac{(\phi a_{11} - a_{12})}{K^3(\det{A})^{3/2}(\phi^2 + \lambda^2 + 2g\lambda\phi)^{1/4}},~~
			U_{1001} = -\frac{\sqrt{\phi^2 + \lambda^2 + 2g\lambda\phi}}{K (\det{A})^{3/2}}(\phi a_{22} + a_{12})
			\\
			U_{1010} = \frac{\sqrt{\lambda^2\phi^2 + 1 - 2g\lambda\phi}}{K (\det{A})^{3/2}}(a_{22} - \phi a_{12}),~~
			U_{1100} = - \frac{a_{12} \sqrt{\lambda}}{K (\det{A})^{3/2}},~~
			U_{1111} =  \frac{(1 - \phi^2) (1 + \frac{3a_{12}^2}{\det{A} })}{(K \det{A})^{3/2}}
			\end{align}
					
			As the corresponding functions in the following integrals are odd:
			\begin{align}
			U_{0001} = U_{0010} = U_{0100} = U_{1000} = U_{0111} =  U_{1011} = U_{1110} = U_{1101} = 0
			\end{align}

\end{document}